\documentclass[sigconf,natbib]{acmart}
\AtBeginDocument{%
  \providecommand\BibTeX{{%
    \normalfont B\kern-0.5em{\scshape i\kern-0.25em b}\kern-0.8em\TeX}}}

\setcopyright{acmcopyright}
\copyrightyear{2024}
\acmYear{2024}
\acmDOI{XXXXXXX.XXXXXXX}

\acmConference[Preprint]{}{2024}{NLM}
\acmPrice{15}
\acmISBN{978-1-4503-XXXX-X/18/06}

\newcommand{\darkblue}[1]{\textcolor[RGB]{17, 87, 78}{#1}}

\usepackage{multirow} 
\usepackage{array}  

\begin{document}

\title{Harnessing PubMed User Query Logs for Post Hoc Explanations of Recommended Similar Articles}

\author{Ashley Shin}
\affiliation{
  \institution{National Library of Medicine}
  \institution{National Institutes of Health}
  \city{Bethesda}
  \state{MD}
  \country{USA}
}
\email{ashley.shin@nih.gov}

\author{Qiao Jin}
\affiliation{
  \institution{National Library of Medicine}
  \institution{National Institutes of Health}
  \city{Bethesda}
  \state{MD}
  \country{USA}
}
\email{qiao.jin@nih.gov}

\author{James Anibal}
\affiliation{
  \institution{Clinical Center}
  \institution{National Institutes of Health}
  \city{Bethesda}
  \state{MD}
  \country{USA}
}
\email{james.anibal@nih.gov}

\author{Zhiyong Lu}
\affiliation{
  \institution{National Library of Medicine}
  \institution{National Institutes of Health}
  \city{Bethesda}
  \state{MD}
  \country{USA}
}
\email{zhiyong.lu@nih.gov}

\renewcommand{\shortauthors}{Shin et al.}

\begin{abstract}
Searching for a related article based on a reference article is an integral part of scientific research. 
PubMed, like many academic search engines, has a ``similar articles'' feature that recommends articles relevant to the current article viewed by a user.
Explaining recommended items can be of great utility to users, particularly in the literature search process. 
With more than a million biomedical papers being published each year, explaining the recommended similar articles would facilitate researchers and clinicians in searching for related articles. 
Nonetheless, the majority of current literature recommendation systems lack explanations for their suggestions.
We employ a post hoc approach to explaining recommendations by identifying relevant tokens in the titles of similar articles.
Our major contribution is building PubCLogs by repurposing 5.6 million pairs of coclicked articles from PubMed's user query logs.
Using our PubCLogs dataset, we train the Highlight Similar Article Title (HSAT), a transformer-based model designed to select the most relevant parts of the title of a similar article, based on the title and abstract of a seed article.
HSAT demonstrates strong performance in our empirical evaluations, achieving an $\mathrm{F_1}$ score of 91.72 percent on the PubCLogs test set, considerably outperforming several baselines including BM25 (70.62), MPNet (67.11), MedCPT (62.22), GPT-3.5 (46.00), and GPT-4 (64.89). Additional evaluations on a separate, manually annotated test set further verifies HSAT's performance. 
Moreover, participants of our user study indicate a preference for HSAT, due to its superior balance between conciseness and comprehensiveness. 
Our study suggests that repurposing user query logs of academic search engines can be a promising way to train state-of-the-art models for explaining literature recommendation.
\end{abstract}



\maketitle

\section{Introduction}

\begin{figure*}[ht!]
  \centering
  \includegraphics[scale=0.65]{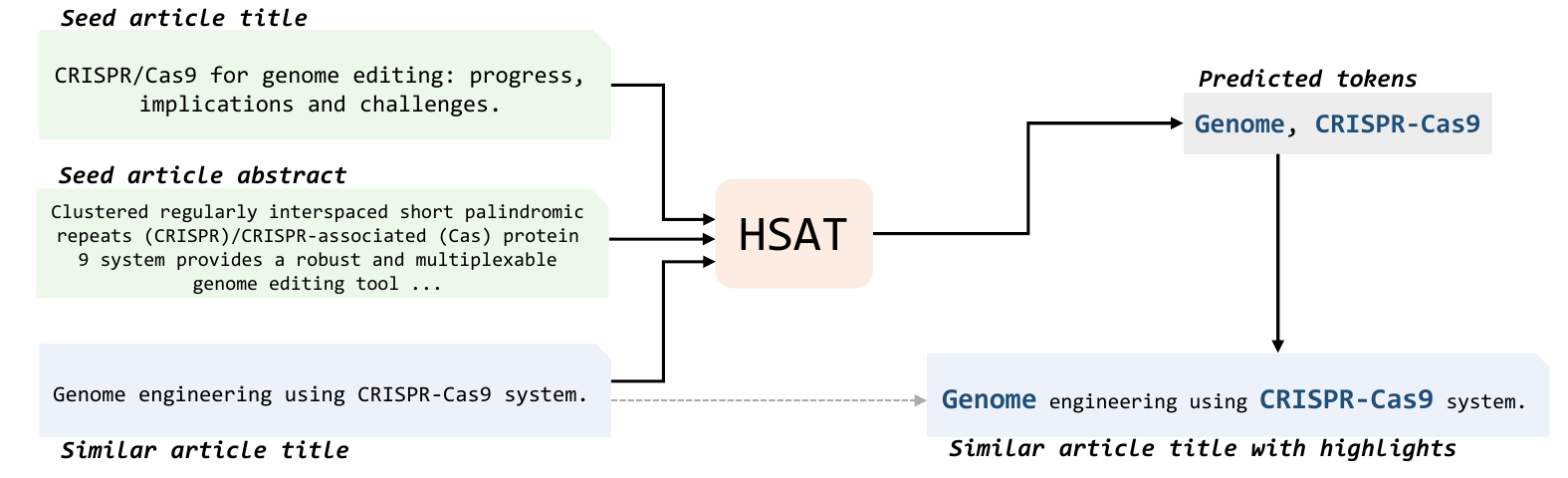}
\caption{Outline of our model, Highlight Similar Article Title (HSAT). Given the title and abstract of a seed article, and the title of a ``similar article,'' HSAT selects the most relevant tokens in the similar article title to be highlighted.}
  \label{fig:overall_diagram}
\end{figure*}

Much of the knowledge gathered through biomedical research is solely accessible via literature \cite{baumgartner2007manual}. 
Literature search, the process of finding scientific articles, is crucial in both patient care and biomedical research \cite{ely2005answering, jin2022state, guido2023screening, gopalakrishnan2019survey}.
However, millions of papers are published each year, making effective literature search increasingly challenging. 
For example, PubMed\footnote{\url{https://pubmed.ncbi.nlm.nih.gov/}}, an academic search engine primarily for biomedical literature, indexes about 36 million articles and adds more than a million yearly.

There are numerous ways to conduct literature search. In addition to the usual query-based searches, a commonly employed approach in literature search is seeking ``similar articles'' based on a reference, or a ``seed article.'' On literature search engines such as PubMed, an article page might display a ``similar articles'' section, where titles and links to papers that the user might also be interested in are shown. 
Based on observations of PubMed user behavior, it is clear that similar articles play a major role in many researchers' information seeking process \cite{Lin2009modeling}.
Users often express a desire to explore other papers related to the same or closely related topics when they come across an article of interest. Essentially, they seek relevant related documents they might not have initially known to search for. 

To meet this user need, for each retrieved article, PubMed displays the top five most closely related articles based on content similarity, with the option to see the full list of similar articles, which might be hundreds of articles long. 
By following links of the recommended similar articles, users can further explore the literature, based on their initial seed articles. 
This feature has not only demonstrated its effectiveness for information seeking but also stands as an integral part of how users interact with PubMed \cite{Lin2009modeling}. 
Furthermore, the same study reveals that once users initiate the exploration of related articles, they tend to persist in this activity, accounting for more than forty percent of their interactions. 
This surpasses the selection of new articles or the initiation of new queries, highlighting the significance of similar article recommendations \cite{Fiorini2018}. 

Given the importance of similar articles in literature search, users naturally would want to see the explanations of similar article recommendations. Inevitably, users need to manually sort through each recommended similar article in order to find articles that fit their needs. Providing explanations as to the ways each particular similar article is relevant to the seed article would greatly facilitate this process, and thus speed up literature search. Articles are recommended for a variety of different reasons, and since users have diverse information needs – for example, one might only be interested in a certain topic, or type of similarity – explaining the recommendations becomes paramount for efficient literature search \cite{jin2023pubmed, zhang2020explainable}.

\begin{figure*}[ht!]
  \centering
  \includegraphics[scale=0.6]{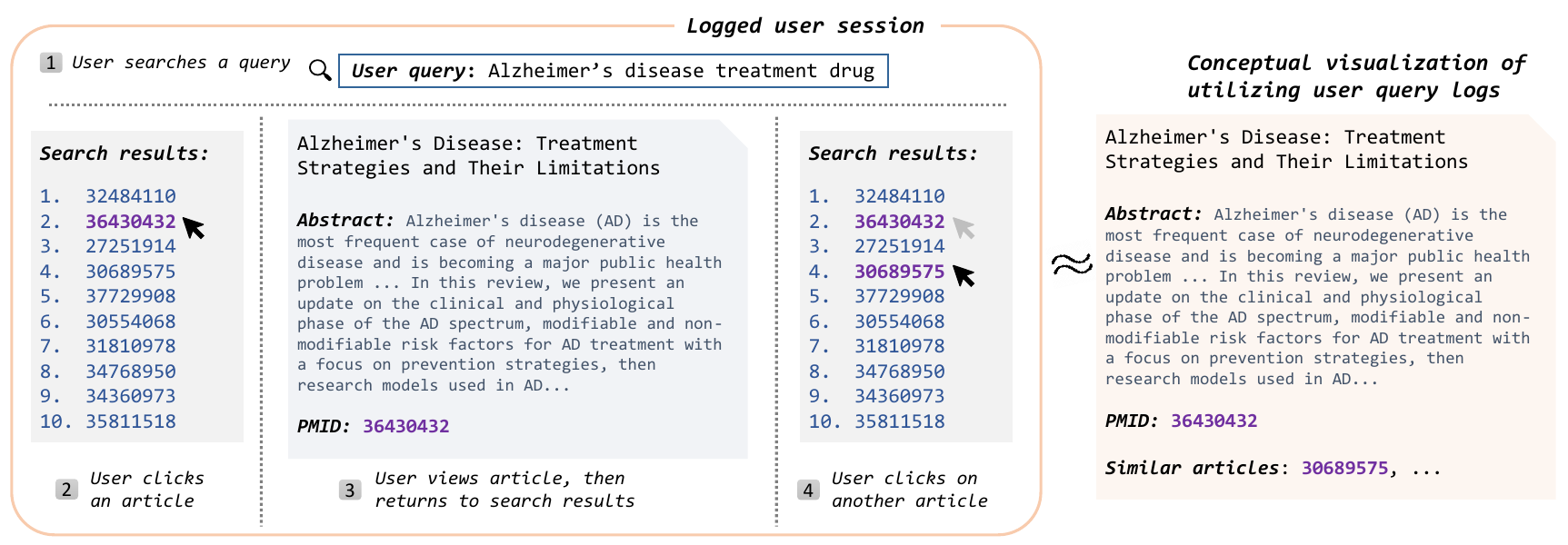}
  \caption{Overview of our novel method to utilize PubMed user query logs for building our dataset, PubCLogs. When a user issues a query and clicks on an article from the initial search results, then views the title and abstract of the clicked article, returns to the search page, and subsequently clicks on another article, we hypothesize that the coclicked articles are likely related, as the user chose the second article after reviewing the content of the first. Thus, pairs of coclicked articles form the foundation for constructing PubCLogs.}
  \label{fig:user_intelligence}
\end{figure*}

One major barrier to providing explanations in academic search engines is data availability. Training a modern NLP model requires large-scale datasets, and to our knowledge, there has not been any such dataset for selecting relevant tokens in similar article titles. To tackle this issue, we utilize the user intelligence from PubMed search logs \cite{Fiorini2018, Dogan2009} to build PubCLogs, a dataset for training and evaluating models for similar article recommendation explanations. We utilize article coclicks as summarized in Figure~\ref{fig:user_intelligence}. We first obtain user coclick data spanning years from 2020 to 2023. The original search logs record a search of a user in a session. We pair up coclicked articles as seed and similar articles. Each pair of articles is linked to a list of corresponding queries that led to the coclicks, and each query's click counts. Using the coclick counts, we select the top queried title tokens to act as the ground truth for each document pair. Our initial collection includes 5.6M such pairs. After filtering, we end up with 1.5M instances for the train set, 47k for the development set, and 47k for the test set. 

Given the importance and challenge of similar article recommendation explanation, in this work, we formulate the explanation task as selecting relevant tokens in the similar article title, given the title and abstract of a seed article. Figure~\ref{fig:overall_diagram} shows our model for the task. We train a transformer-based sequence tagging model, Highlight Similar Article Title (HSAT), using the PubCLogs dataset we construct with article coclicks from PubMed user query logs\cite{transformer}. We show that it vastly outperforms common baselines such as BM25 \cite{bm25} and SBERT \cite{mpnet}, as well as GPT-3.5 \cite{chatgpt} and GPT-4 \cite{gpt4} on independent holdout test set portion of PubCLogs. 

On the test set, HSAT achieves a $\mathrm{F_1}$ of 91.72 percent, outperforming baselines such as BM25 (70.62), MPNet (67.11), MedCPT (62.22), GPT-3.5 (46.00), and GPT-4 (64.89). We also create a smaller, human-annotated test set to further verify our results. On the manually annotated test set, HSAT achieves a $\mathrm{F_1}$ of 80.62, again outscoring models such as BM25 (57.21), Word2Vec (54.81), GPT-3.5 (42.06), and GPT-4 (68.31). Our evaluation also includes case studies and user studies. Our A/B user studies show that our two annotators prefer HSAT outputs over GPT-4 outputs, 55 to 38 and 59 to 31. Our evaluations indicate the utility of our PubCLogs dataset, which we demonstrate by training HSAT with a binary sequence tagging objective that outperforms common baselines. HSAT performance suggests that utilizing user query logs is a promising approach to explaining article recommendations.

In summary, our contributions are three-fold:

\begin{itemize}
    \item We formulate the task of explaining similar article recommendations as selecting tokens in the similar article title that are the most relevant in relation to the seed article. 
    \item We construct a dataset, PubCLogs, using a novel method to repurpose the user search logs. We consider coclicked articles as seed and similar article pairs, and utilize queries that correspond to the coclicks to build the ground truth of relevant tokens in the similar article title. 
    \item Utilizing the PubCLogs dataset, we train Highlight Similar Article Title (HSAT), our transformer-based model, on a sequence tagging objective. HSAT outperforms baseline comparisons such as BM25, Word2Vec, GPT-3.5, and GPT-4 on our evaluations, which consist of the holdout test set of PubCLogs, a separate manually labeled test set, and a user preference study. 
\end{itemize}

\section{Related work}
\subsection{Post hoc model of explainability}
There are two major approaches to recommendation explanation: one can seek explainability in the recommendation methods or the results \cite{zhang2020explainable}. 
The model-intrinsic approach aims to develop interpretable models that naturally lend themselves to enhanced transparency and explainability \cite{zhang2014explicit}. 
In contrast, the model-agnostic, or ``post hoc,'' approach aims to provide explanations independently from the model that created the recommendations \cite{peake2018posthoc, singh2019explainable}. 
Since the explaining model and the recommending models are separate, the recommender system need not be interpretable, which offers more flexibility in model development given the black-box nature of many modern deep learning models \cite{jain2019attention}. 
Though explainability and interpretability are frequently conflated, the post hoc approach illustrates that interpretability is but an approach to explainability \cite{qingyao2021model-agnostic}.

\subsection{Production academic search engines}
Despite the clear utility of explaining recommendations to users, most academic search engines are not sufficiently explainable. 
In its ``related articles'' section, Semantic Scholar \footnote{https://www.semanticscholar.org/} provides short summary snippets of each article called ``TLDR'' \cite{cachola2020TLDR}. 
However, as of January 2024, it does not highlight any key words in the title or TLDR of each recommended similar article, which means users have to read the entire TLDR, which can be relatively more time-consuming. 
Likewise, while Google Scholar\footnote{https://scholar.google.com/} does provide the first few lines of the abstract for its related articles, it does not do any highlighting or further annotations. 
Similarly, PubMed does not bold or highlight any part of the similar articles' titles in its similar articles section \cite{Lin2007PubMedRA}. 
We argue that highlighting relevant words in the similar article title is a practical and effective first step towards article recommendation explanations.

\subsection{Sequence tagging}
We initially considered training with an extractive QA task, where the ``question'' would be the context, i.e., the title and abstract of the seed article, and the ``answer'' would be the parts of the similar article title to be highlighted. But we ultimately choose sequence tagging because since the tokens to be highlighted would often not be contiguous, our task more closely resembles a multi-span extractive QA task. In most preexisting extractive QA benchmarks such as SQuAD \cite{SQuAD}, SQuAD2.0 \cite{SQuAD2.0}, QuAC \cite{QuAC}, and HotpotQA \cite{HotpotQA}, models are to return a single, contiguous span of answer from each given context. 

Multi-span extractive QA models can be more complicated to implement than their single-span counterparts, but as Setal et al. \cite{segal-2020-simple} show, a simple way to implement multi-span extraction is by treating it as a sequence tagging task. In the more recent MultiSpanQA benchmark \cite{MultiSpanQA}, the sequence tagging approach vastly outperforms single span models as expected. Despite the simplicity, sequence tagging achieves SoTA results on span-extraction questions from DROP \cite{dua2019drop} and QuoREF \cite{dasigi2019quoref}. The sequence tagging approach only requires finetuning one encoder model. Thus, we consider the sequence tagging approach a good balance between performance and simplicity in implementation.

\begin{figure*}
  \centering
  \includegraphics[scale=0.6]{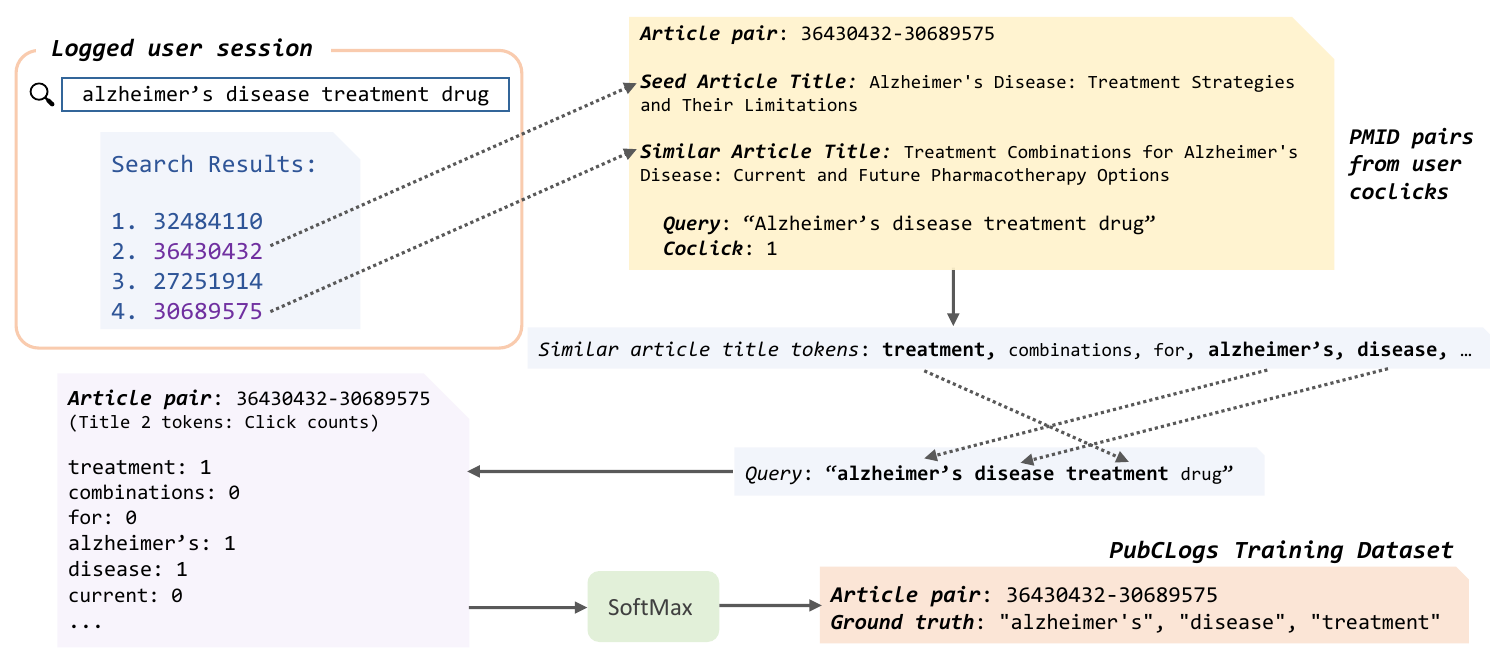}
  \caption{Overview of the PubCLogs dataset construction process: for each coclicked article pair, the initial article represents the seed article, and the related article clicked subsequently represents the similar article. For each token in the title of the similar article, we aggregate the number of coclicks from queries that included the title token. We apply a softmax function to these click counts and establish a predefined threshold, \textit{P}, to identify the most frequently queried similar article title tokens, which are then used as the ground truth labels for the PMID pair.}
  \label{fig:data_collection}
\end{figure*}

\section{Method}
In this section, we will outline the preprocessing of user query logs and the construction of PubCLogs, our dataset, as well as the training process of our model, HSAT.

\subsection{Building PubCLogs}

\paragraph{\textbf{Preprocessing}}
Figure~\ref{fig:data_collection} summarizes our overall process for data collection and dataset construction. In building PubCLogs from PubMed user query logs, our key insight is in the interpretation of coclicked articles as pairs of seed and similar articles, a concept depicted in Figure~\ref{fig:user_intelligence}. We start with user coclick data spanning the years 2020 to 2023. The data logs each user's search within a session. Specifically, when a user executes a query and selects two or more articles from the results page, this action is recognized as a coclick instance for the article pair, e.g., the PMID of a seed article (PMID1) and the PMID of a similar article (PMID2), where PMID1 is the article that appears higher in the results page. PMID1 would be considered to be the ``seed article'' and PMID2 is treated as the ``similar article.'' In our starting material, each pair of articles, PMID1-PMID2, is linked to a dictionary of queries and their corresponding click counts, where the number of clicks is shown for each unique query. For each coclick instance, we then increment the coclick count corresponding to a unique query that retrieved the PMID pair.  

From there, we tokenize the queries using the standard NLTK tokenizer \cite{nltk2009}. Each token in the query is multiplied by the click count for the query. And we combine the coclick counts for each unique token in the list of queries to get a combined dictionary that links each unique token to the total coclick count, for each unique pair PMID1-PMID2. We apply softmax to the click counts and use a threshold \textit{P} to select the top queried title tokens to act as the ground truth for the document pair. 
The resulting PubCLogs entry is a set of unique tokens in the similar article title corresponding to each seed and similar article pairs.
Our initial collection includes 5.6M article pair entries.
We then filter out entries with less than 20 combined clicks, article titles less than seven tokens long, and less than 3 title tokens with nonzero click counts to reduce the noisy long tail. 

\paragraph{\textbf{Dataset analysis}}
After preprocessing and filtering, 1.5M article pair entries remain for the train set, 47k for the development set, and 47k for the test set. The average title length is 17.5 tokens. 

Our dataset before filtering for combined clicks is 4.3GB. When we filter for instances with over 20 combined clicks, it is 1.2GB, 300MB for 50 clicks, and 100MB for 100 clicks, so we see that click counts follow a power law. 

\begin{table*}[ht!]
    \centering
    \begin{tabular}{l|ccccccc|ccccccc}
        \toprule
        Model & \multicolumn{7}{c|}{PubCLogs Test Set} & \multicolumn{7}{c}{Manually Annotated Test Set} \\
        \cmidrule{2-15} & \multicolumn{3}{c|}{Token-level}  & \multicolumn{4}{c|}{Title-level} & \multicolumn{3}{c|}{Token-level} &  \multicolumn{4}{c}{Title-level} \\
        \cmidrule{2-15} & R  & P  & $\mathrm{F_1}$ & R & P & $\mathrm{F_1}$ & L & R  & P  & $\mathrm{F_1}$  & R  & P & $\mathrm{F_1}$ & L \\
        \midrule
        HighlightAll & \textbf{100.0} & 21.45 & 35.32 & \textbf{100.0} & 21.45 & 35.05 & 17.5 & \textbf{100.0} & 21.49 & 35.35 & \textbf{100.0} & 21.49 & 35.10 & 17.5\\
        Overlapper & 61.91 & 24.71 & 35.32 & 59.43 & 24.27 & 34.47 & 9.2 & 67.75 & 25.04 & 36.57 & 65.01 & 24.27 & 35.35 & 9.8 \\
        BM25 \cite{bm25} & 67.28 & 74.30 & 70.62 & 64.56 & 71.29 & 67.76 & 3.2 & 55.53 & 59.00 & 57.21 & 53.43 & 56.13 & 54.75 & 3.2 \\ 
        \midrule
        Word2Vec \cite{word2vec} & 48.57 & 52.26 & 50.35 & 46.51 & 50.09 & 48.23 & 3.2 & 55.65 & 54.00 & 54.81 & 53.33 & 51.08 & 52.18 & 3.2 \\
        BioWord2Vec \cite{bioword2vec} & 39.06 & 41.78 & 40.37 & 37.45 & 40.10 & 38.73 & 3.2 & 50.50 & 49.00 & 49.74 & 48.68 & 47.08 & 47.87 & 3.2  \\
        MPNet \cite{mpnet} & 64.83 & 69.56 & 67.11 & 62.27 & 66.76 & 64.43 & 3.2 & 73.78 & 74.33 & 74.06 & 70.90 & 70.48 & 70.69 & 3.2\\
        MedCPT \cite{medcpt} & 60.01 & 64.59 & 62.22 & 57.68 & 61.08 & 59.33 & 3.3 & 59.52 & 61.67 & 60.57 & 57.13 & 57.28 & 57.20 & 3.4 \\
        \midrule
        GPT-3.5-turbo \cite{chatgpt} & 51.75 & 41.40 & 46.00 & 49.50 & 42.43 & 45.69 & 3.8 & 47.08 & 38.00 & 42.06 & 45.26 & 41.05 & 43.05 & 3.5 \\
        GPT-4 \cite{gpt4} & 75.68 & 56.79 & 64.89 & 72.59 & 58.25 & 64.63 & 4.4 & 80.47 & 60.49 & 69.06 & 77.05 & 59.57 & 67.19 & 4.3  \\ 
        \midrule
        HSAT (ours) & 91.50 & \textbf{92.15} & \textbf{91.83} & 87.77 & \textbf{88.69} & \textbf{88.23} & 3.6 & 79.97 & \textbf{81.28} & \textbf{80.62} & 76.34 & \textbf{78.56} & \textbf{77.44 } & 3.4 \\  
        \bottomrule
    \end{tabular}
    \caption{Results of evaluations on our test dataset. We compare standard baseline models versus our model, Highlight Similar Article Title (HSAT). We differentiate token-level and title-level metrics because titles can contain multiple instances of unique tokens. Best performances are bolded. L refers to the average length of predicted outputs, in terms of tokens.}
    \label{tab: Main results}
\end{table*}

\subsection{Training HSAT}
\paragraph{\textbf{Setup}}
We initialize a BERT-based model with PubMedBERT weights \cite{pubmedbert} and train it on a sequence tagging task using PubCLogs. 
We use Pytorch \cite{pytorch} and Hugging Face Transformer libraries \cite{huggingface} for training and inference, and train with Adam \cite{adam}. We use two labels for sequence tagging: 0 for nonrelevant title tokens and the context, i.e. the title and abstracts for the seed article; and 1 for relevant title tokens, i.e. tokens to be highlighted. After a grid search of \texttt{Learning Rate} $\in \{2e^{-5}, 3e^{-5}, 4e^{-5}, 5e^{-5}, 6e^{-5}\}$ and $\beta_1 \in \{0.9, 0.99, 0.999\}$, we choose a learning rate of 5e-5 and $\beta_1$ of 0.9 for the Adam optimizer. We find that a cosine learning rate scheduler with warm up performs better than a linear one. We use a cosine learning rate scheduler with 20k warm up steps and train for 10 epochs without decay. The batch size 64. We otherwise use the Pytorch and Adam default hyperparameters for finetuning BERT. Training takes about 70 hours on two Nvidia Tesla V100-SXM2 32GB GPUs. On the development set, we find that the best performing checkpoint is one at 180,000 training steps. This checkpoint is the HSAT model that we refer to throughout this work, including in evaluations. 

For inputs to our model, we use the abstract and title of the seed article, the title of the similar article, and similar article tokens to be highlighted. We tokenize all with the PubMedBERT tokenizer before concatenating them. HSAT inputs are formatted like so:
\[\texttt{Tokens:} \texttt{[CLS]} \{t_1, t_2, ..., t_n \} \{a_1, a_2, ...\} \texttt{[SEP]} \{t'_1, t'_2, ..., t'_m\}\] 
\[\texttt{Labels:} \text{[1 for relevant } t'_i\text{; 0 for all } t_i, a_i\text{, and nonrelevant } t'_i\text{]}\]
${t_i}$ denotes the tokenized seed article title, $a_i$ denotes the tokenized seed article abstract, and $t'_i$ denotes the tokenized similar article title. For the similar article, we choose to only use the title because of the length limit on BERT architectures, and to better simulate users' behavior when they encounter similar articles, since only the titles are displayed on the search engine due to space constraints.

We noticed a more than a 10 percent points absolute increase in $\mathrm{F_1}$ after adding segment embeddings to better differentiate between the seed and similar articles in the concatenated inputs. Since we start with PubMedBERT, we also train a linear classifier in order to convert each token embedding into labels, relevant or nonrelevant. We take the last hidden state of HSAT, and use the trained linear classifier to label each similar title token.

\paragraph{\textbf{Inference}}
For inference, the model outputs binary labels, given concatenated tokens of PMID1 abstract and title and PMID2 title. We match the tokens with labels, and tokens with ``1'' are tokens to be predicted. Each model output is thus a subset of the tokenized title of the similar article.
Moreover, because HSAT outputs are based on the PubMedBERT tokenizer, the tokenization of the same similar article titles will be different from the other baseline models, which is fed input tokenized by NLTK tokenizer \cite{nltk2009}. Thus, we apply postprocessing to convert title tokens tokenized by BERT into NLTK-style tokens. 
We must also distinguish title and token level inference, which is shown well in Table~\ref{tab: Main results}. In token-level inference, we stop after the postprocessing. For title-level inference, we match all tokens in the similar article title that match each token in our predicted output set.

\section{Evaluation}
We evaluate HSAT on the test set of PubCLogs and a manually annotated test set (``manual set''). We also conduct user studies to further confirm that users prefer HSAT outputs over those from GPT-4.

\begin{table*}
    \centering
    \begin{tabular}{l|ccc|ccc|ccc|ccc}
        \toprule
        & \multicolumn{3}{c|}{Top 0.1\%} & \multicolumn{3}{c|}{Top third} & \multicolumn{3}{c|}{Middle third} & \multicolumn{3}{c}{Bottom third} \\
        \cmidrule{2-13}
        Model & R  & P  & $\mathrm{F_1}$ & R  & P & $\mathrm{F_1}$& R  & P  & $\mathrm{F_1}$& R & P  & $\mathrm{F_1}$ \\
        \midrule
        HighlightAll & \textbf{100.0} & 24.37 & 39.19 &  \textbf{100.0} & 21.88 & 35.90 & \textbf{100.0} & 22.06 & 36.15 & \textbf{100.00} & 20.40 & 33.88    \\
        Overlapper & 69.65 & 33.47 & 45.21 &  63.75 & 25.75 & 36.68 & 61.73 & 25.00 & 35.59 & 60.23 & 23.37 & 33.68     \\
        BM25 \cite{bm25} & 69.57 & 61.70 & 65.40  & 69.79 & 70.26 & 70.02 & 69.04 & 75.55 & 72.14 & 63.00 & 77.11 & 69.34    \\
        \midrule
        Word2Vec \cite{word2vec} & 67.84 & 56.03 & 61.37 & 54.56 & 53.56 & 54.06 & 50.39 & 54.27 & 52.26 & 40.72 & 48.94 & 44.45     \\
        BioWord2Vec \cite{bioword2vec} & 62.09 & 51.77 & 56.47 & 43.88 & 42.67 & 43.27 & 40.28 & 43.07 & 41.63 & 32.99& 39.60 & 36.00     \\
        MPNet \cite{mpnet} & 83.16 & 68.09 & 74.87 &  71.62 & 69.56 & 70.58 & 66.18 & 70.90 & 68.46 & 56.68 & 68.20 & 61.91    \\
        MedCPT \cite{medcpt} & 72.62 & 58.87 & 65.02 & 65.69 & 64.00 & 64.83 & 61.34 & 65.92 & 63.54 & 52.98 & 63.86 & 57.91   \\
        \midrule
        GPT-3.5 \cite{chatgpt} & 51.38 & 35.46 & 41.96 & 56.71 & 41.53 & 47.95 & 53.73 & 42.89 & 47.70 & 44.80 & 39.76 & 42.13    \\
        GPT-4 \cite{gpt4} & 87.13 & 55.11 & 67.51 & 80.05 & 56.27 & 66.08 & 77.44 & 58.62 & 66.73 & 69.54 & 55.48 & 61.72   \\
        \midrule
        HSAT (ours) & 97.62 & \textbf{97.48} & \textbf{97.55} & 92.57 & \textbf{93.12} & \textbf{92.84} & 92.18 & \textbf{93.09}& \textbf{92.64} & 89.38 & \textbf{90.34} & \textbf{89.86}  \\
        \bottomrule
    \end{tabular}
    \caption{Evaluations on subsets of the PubCLogs test set. Subsets are grouped by combined coclick counts of article pairs. We compare standard baseline models versus Highlight Similar Article Title (HSAT).}
    \label{tab:extended_model_comparison}
\end{table*}

\subsection{Baselines}
We compare the performance of our HSAT model with the following baselines: HighlightAll, Overlapper, BM25, Word2Vec, BioWord2Vec, MPNet, and MedCPT. 
The results on the PubCLogs test set are shown in Table 1. HighlightAll and Overlapper are the trivial baselines. HighlightAll selects all possible tokens, hence the recall of 100\% and low precision. Overlapper selects all similar article title tokens that are also in the seed article title. BM25 is implemented following the standard equation \cite{bm25}.   
We opt not to use any single-span extractive QA models because as the MultiSpanQA evaluations \cite{MultiSpanQA} show, single-span models are greatly outperformed by the sequence tagging baseline, sometimes by over 40 $\mathrm{F_1}$ points. 

Here we give more details on the implementations of our baselines. There were no straightforward ways to implement baselines. For the most ``accurate'' comparisons, we would ideally consider token binary classification models. However, most sequence tagging models follow the common \texttt{BIO} scheme, which is not suitable for our use case. For example, \texttt{biomedical-ner} is a BERT-based sequence tagging model that recognizes biomedical entities such as biological structures and symptoms \cite{raza2022large}. Since our goal is to select tokens based on their relevance with respect to a span, rather than labeling each token, \texttt{biomedical-ner} would not be suitable to serve as a baseline for our task. 

Thus, we end up implementing the following baselines – with the exception of HighlightAll and Overlapper – by calculating the relevance scores for each token in similar article title, ranking them by the relevance score, and choosing the top 3 similar article title tokens. Calculating the relevance score for each baseline depends on whether the model is more suited to process tokens or spans. We directly compute the relevance between a token and the entire span of seed article title when the model is designed to process a span, such as BM25, MPNet, and MedCPT. 
We denote the seed article title as $T$, where $T = {t_1, t_2, ... t_n}$. The similar article title is $T'$, where $T' = {t'_1, t'_2, ... t'_n}$. We can represent the similarity score of a similar article token and the seed title as: \[rel(t'_i, T)\]
But for models that only work best at token-level, such as Word2Vec and BioWord2Vec, we can compute the relevance score of a similar article token by: 
\[rel(t'_i, T) = \sum_{j=1}^{n} \text{rel}(t'_i, t_j)\]

The metrics mentioned above are assuming token-level evaluation. Table~\ref{tab: Main results} contains both token-level and title-level evaluations. The only difference between them is that title-level accounts for duplicate tokens in titles.

\paragraph{\textbf{HighlightAll.}} As the name indicates, this baseline simply selects every token of similar article title as relevant. As expected, it achieves a recall of 100 but has low precision.
\paragraph{\textbf{Overlapper}} Here, we select all tokens in similar article title that are also in seed article title. We use a list of stopwords and an index of inverse document frequency to filter out some tokens. This is a decent commonsense baseline because any token that is also present in seed article title is bound to be relevant, but it inevitably cannot perform semantic matching.

\paragraph{\textbf{BM25}} We implement BM25 using a standard formula \cite{bm25}, with a $k_1$ of 0.5 and $b$ of 0.3. 

\paragraph{\textbf{Word2Vec and BioWord2Vec}} The Word2Vec model is implemented using implementations from FastText \cite{word2vec}. The BioWord2Vec model \cite{bioword2vec} we use is from the NCBI and trained on PubMed and MIMIC-III clinical notes. 

\paragraph{\textbf{MPNet}} MPNet \cite{mpnet} is a model designed to overcome the drawbacks of BERT \cite{bert} and XLNet \cite{xlnet}. The masked language modeling objective has the drawback of the pretrain-finetune descrepancy, since the trained model will likely not encounter artificial tokens such as \texttt{[MASK]}. XLNet addresses these pitfalls with an autoregressive permutation language modeling objective. MPNet improves on both BERT and XLNet by introducing the masked and permuted language modeling objective. We use the MPNet model from the Sentence-Transformers library \cite{mpnet}.

\paragraph{\textbf{MedCPT}} The MedCPT retriever is implemented using a bi-encoder architecture comprised of a query encoder and a document encoder. The bi-encoder was designed to calculate the relevance of a document to a query by computing the cosine similarity of embeddings from the respective encoders. Here, we adapt it to our use case by using the query encoder to encode $token2_i$ and the document encoder to encode $title1$, since the query encoder is designed for shorter inputs and the document encoder is designed for longer inputs.

\paragraph{\textbf{GPT-3.5 and GPT-4}} We use the OpenAI API to conduct experiments with GPT-3.5 (\texttt{gpt-3.5-turbo}) and GPT-4 (\texttt{gpt-4}), through Microsoft Azure's OpenAI services. Inference temperature is set to 0 so that the outputs are as deterministic as possible. The GPT models are given the title and abstract of the seed article and the title of the similar article and asked to return the top 4 most relevant tokens from the similar article title, based on the seed article. 

An interesting problem is deciding the cutoff point of the ranked list of $token2_i$ to include as the relevant tokens to highlight. We experiment with two major methods: using softmax and threshold P, and selecting the top K tokens. Softmax yields slightly better results, but only after additional manually hardcoding. For example, if 7 tokens pass the threshold, and the length of similar article title is 10 tokens, then we only take the top 4 tokens. That is because even after the Softmax, the scores generated from baseline models are not sufficiently differentiable, so a threshold either selects too many or too few tokens most of the time. Thus, we opt to go with top K tokens to implement our baselines -- top 3 for most baseline models and top 4 for GPT-3 and GPT-4. 

\begin{figure}
  \centering
  \includegraphics[scale=0.56]{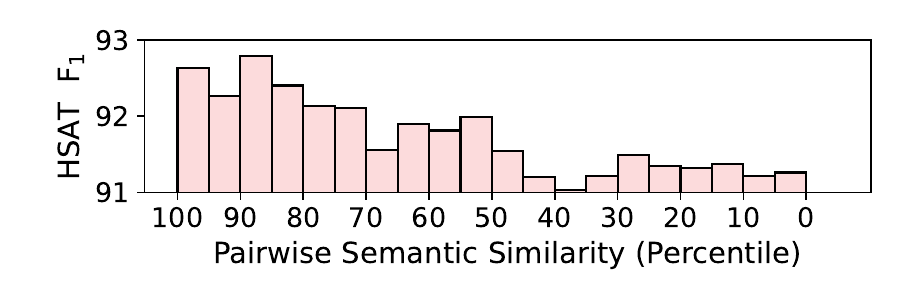}
\caption{$\mathrm{F_1}$ scores of our model, Highlight Similar Article Title (HSAT) on subsets of the PubCLogs test set, divided based on article pair semantic similarity using MedCPT.}
  \label{fig:distribution_semantic_similarity}
\end{figure}

\subsection{Manually annotated test set}
In order to further verify our evaluation results on PubCLogs, we create a separate manually annotated test set of 100 samples. From the top 1000 most highly clicked articles from our search logs, we randomly choose 20 articles to act as our seed articles. For each of those 20, we use NCBI Entrez Programming Utilities (E-Utilities) \cite{EntrezUtilities} to retrieve the top 20 similar articles, from which we randomly select five. Our annotator labels the 100 seed and similar article pairs by selecting the most relevant tokens from the similar article title. The righthand side of Table~\ref{tab: Main results} shows our results on the manual set. The results, similarly to PubCLogs results, indicate that HSAT indeed outperforms all the baseline models evaluated.

\subsection{Results}

We choose standard metrics for evaluation: recall, precision, and $\mathrm{F_1}$. For each instance in the test set, i.e. the title and abstract of a seed article and the title of a corresponding similar article, there is a list of tokens to be highlighted (``true tokens''). Each baseline model and our model, HSAT, outputs a list of tokens to be highlighted (``predicted tokens''). When calculating our metrics, relevant elements are the true tokens and the retrieved elements are the predicted tokens. 

Table~\ref{tab: Main results} shows the main evaluation results on our test dataset, which is a holdout set from our dataset, PubCLogs, constructed from PubMed coclick user logs. Table 2 is our evaluation on a manually labeled test set. In Table~\ref{tab: Main results}, HighlightAll performs best on recall as expected. The precision is around 21, indicating that PubCLogs's ground truth deems about one-fifth of all title tokens as relevant. The ideal percentage of relevant tokens is subjective, of course, but we choose a relatively lower number to focus users' attention on the most important and relevant topics the recommendations represent. The Overlapper is our commonsense baseline, since we would naturally expect that words that appear in both seed and similar article titles would be the ones on the mutually shared topics. Both HighlightAll and Overlapper have a $\mathrm{F_1}$ of 35. BM25 \cite{bm25} performs relatively well, at a precision of 74.30 and $\mathrm{F_1}$ of 70.62. As many recent studies continue to show, BM25 remains a strong baseline for many retrieval and NLP tasks, including article recommendation explanation \cite{Lin2019Hype}. 

Interestingly, for N-gram models such as Word2Vec \cite{word2vec}, the general-purpose Word2Vec outperforms BioWord2Vec \cite{bioword2vec}, at $\mathrm{F_1}$ of 50.35 versus 40.37. However, since these models were not explicitly trained for this task, we can expect some noise and variance in evaluation, especially since they both perform worse than BM25. The MPNet \cite{mpnet} and MedCPT \cite{medcpt} fare much better, at $\mathrm{F_1}$ of 67.11 and 62.22. 

Our most sophisticated baselines are the two GPT models from OpenAI, GPT-3.5 \cite{chatgpt} and GPT-4 \cite{gpt4}. Although over many magnitudes larger in size and inference cost compared to HSAT, GPT-3.5 does quite poorly, with a $\mathrm{F_1}$ of 46.00. GPT-4 does much better, at 64.89, but still underperforms BM25 by over 5.73 percentage points. Interestingly, although prompted to only return the top 4 most relevant tokens, GPT-4 sometimes returns over 5 tokens, which is why the recall is pretty high at 75.68. And due to the excessive number of selected tokens, the precision is lower at 56.79. Overall, the GPT models fare poorly, which is to be expected given the recent studies exposing the limits of LLMs, especially in more specialized domains such as science and medicine. 

HSAT holds the best precision at 92.29 and $\mathrm{F_1}$ at 91.72, outperforming GPT-4 by 26.83 percentage points. GPT-4 does perform better than GPT-3.5, but it still lags other baselines such as BM25 and MPNet. For example, on the PubCLogs test set, GPT-4 has a $\mathrm{F_1}$ of 64.89, which is lower than BM25's 70.62 and MPNet's 67.11.

Our main evaluations are in token-level, but we also show title-level metrics in Table~\ref{tab: Main results}. Performances on PubCLogs test set for title-level is slightly lower compared to token-level across the board, perhaps due to the added complexity of duplicate tokens in titles.

\begin{figure}
  \centering
  \includegraphics[scale=0.5]{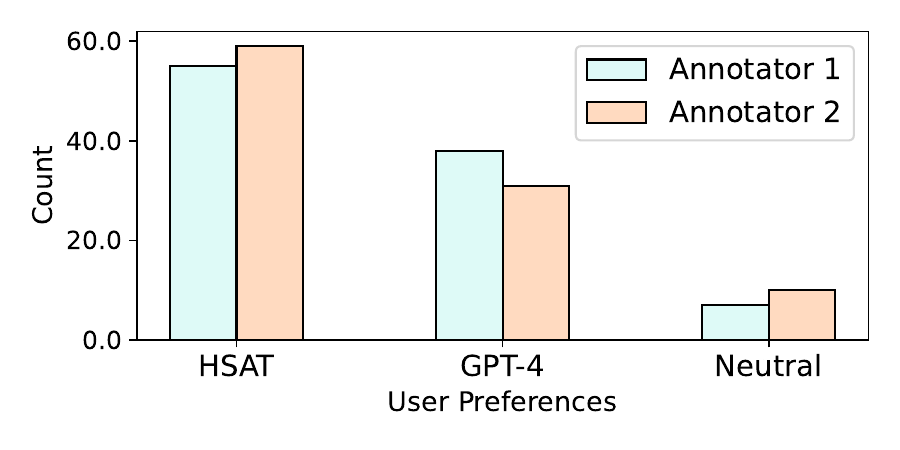}
    \caption{Results of our user preference studies of HSAT outputs versus GPT-4 outputs. The neutral column indicates that both outputs were similar in quality.}
    \label{fig:user_studies}
\end{figure}

\subsection{User studies}
We conduct a user study with two annotators with graduate-level biomedical knowledge to determine whether users would prefer recommendation explanations from HSAT or GPT-4. We use HSAT and GPT-4 outputs from our manually annotated set, which has a size of 100. We randomize the order of the two outputs shown for each entry, and ask each study participant to rank the two. The manual set was ranked by the participants. Our user study shows that both users significantly favor HSAT outputs over GPT-4. Figure~\ref{fig:user_studies} shows our user study results. Annotator 1 preferred HSAT over GPT-4 55 to 38, while annotator 2 preferred HSAT by 59 to 31. The annotator agreement was 89\%. The annotators agreed that HSAT tends to be more succinct in its answers, while highlighting topics that both seed and similar articles share.

\subsection{Case studies}
We show three case studies in Table~\ref{tab:case_studies}. Two entries are from the test set of PubCLogs and one is from our manually-annotated test set. 
The title and abstract for the seed article and the title of the similar article are shown on the leftmost column. 
Each row representing the model has a corresponding similar article title. 
Model names are in the middle column, and the similar article titles are shown in the rightmost column, where the predicted tokens are highlighted. 

In the ``Ground truth'' rows, we show the ground truths from PubCLogs training set that we trained HSAT on. The ground truth tokens are selected by processing queries that led to coclicked article pairs. In the HSAT rows, outputs from the model are displayed. In the first entry, with PMID pair 33332292 and 33301246, HighlightAll highlights all tokens in the similar article title (``similar title''), which leads to perfect recall of 100.0 but low precision. Next is Overlapper, which selects tokens that are in both seed and similar titles. The Overlapper is a good, commonsense baseline. Intuitively, articles with titles that share many terms will tend to be similar. Overlapper works best when the similar title tokens lexically match tokens in the seed title. But due to its nature, it will inevitably also select tokens that are of low value to users, such as stopwords. And more importantly, it will miss tokens that are semantically, but not lexically matched with a seed title token. 

BM25 is a strong, standard baseline. In the first entry, however, we see that it selects ``and'' and ``the,'' which are stopwords that do not provide insight into the recommendations. 
This is likely due to the fact that the model was forced to return the top three tokens, even in entries where there were no clear top tokens. 

Word2Vec and BioWord2Vec miss critical information in the first entry. Both the seed and similar article titles mention COVID-19, so it seems that both ``COVID-19'' and ``vaccine'' are relevant tokens. However, Word2Vec and BioWord2Vec fail to highlight ``Covid-19'' and select ``safety'' and ``efficacy'' instead, which are lower-value words. BioWord2Vec outputs show a similar pattern in the second entry, where it only selects ``diet'' but not ``ketogenic,'' which appears in both seed and similar titles. 
In the third entry, Word2Vec selects ``arthroplasty,'' which might be an important topic in the similar article on its own. However, the seed article seems to be about turmeric extracts and curcumin, and neither the title nor the abstract mentions arthroplasty at all. Since our goal is to select information that is relevant in relation to the seed article, ideal outputs should not include such words.

MPNet performs well despite being a general domain model. In the third entry, its outputs exactly match the ground truth, and produce acceptable results in the first two as well. It also shows relatively consistent performances in the PubCLogs test set and the manually annotated test set, showcasing its robustness. MedCPT retriever does not do as well, but still shows decent performance. It makes a critical mistake in the first entry by selecting ``and'' but not ``Covid-19.'' MedCPT's lower performance may be due to the fact that the query encoder and document encoders of MedCPT were trained to encode text much longer than our usage. 

As expected, GPT-4 does much better than GPT-3.5. GPT-3.5 makes trivial mistakes such as highlighting ``safety'' and ``efficacy'' but missing ``Covid-19.'' And although GPT-4 does better than GPT-3.5, it still underperforms other baselines such as BM25 and MPNet. For example, on the PubCLogs test set, GPT-4 has a $\mathrm{F_1}$ of 64.89, which is lower than BM25's 70.62 and MPNet's 67.11. Although we can argue that the lower performance is due to the fact that GPT models are general domain, so are BM25 and MPNet. Given the nonstellar performance and exorbitant training and inference cost of the GPT models, it seems for now that LLMs are not practical for large-scale usage in article recommendation explanation.

In the second entry, both the seed and similar articles are comparing ketogenic diets and low-fat diets for weight loss. Although almost all of our baselines correctly select ``ketogenic diet'' and ``low-fat diet,'' none of them highlight ``weight,'' which is arguably an important commonality between the articles. In this sense, HSAT correctly includes all three phrases in its output. In the third entry, ``rheumatoid'' is part of the similar title but is not mentioned in the seed article, yet is part of the PubCLogs ground truth and outputs from HSAT, and almost all baselines including Word2Vec, MPNet, GPT-3.5, and GPT-4. While this is a critical error, it is one made by most of the baselines as well. Interestingly, Overlapper and BM25 do not make this error, showing that each model has its strengths and weaknesses, and a fusion model that combines multiple model outputs is likely to be better than any one model on its own.

\begin{table*}
    \centering
    \begin{tabular}{|m{4.5cm}|@{\hskip 0.1pt}|m{2cm}|m{10cm}|} 
        \hline
        \multirow{3}{4.5cm}{\newline \textbf{Seed article PMID }: 33332292 \newline \textbf{Similar article PMID}: 33301246 \newline\newline \textbf{Seed article title}: The Advisory Committee on Immunization Practices' Interim Recommendation for Use of Pfizer-BioNTech COVID-19 Vaccine - United States, December 2020.} & \textbf{Similar article title} & Safety and Efficacy of the BNT162b2 mRNA Covid-19 Vaccine. \\ \cline{2-3}
        & \textbf{Ground truth} & Safety and Efficacy of the \textbf{\darkblue{BNT162b2}} mRNA \textbf{\darkblue{Covid-19}} \textbf{\darkblue{Vaccine}}. \\ \cline{2-3}
        & \textbf{HSAT (ours)} & Safety and Efficacy of the \textbf{\darkblue{BNT162b2}} mRNA \textbf{\darkblue{Covid-19}} \textbf{\darkblue{Vaccine}}. \\ \cline{2-3}
        & HighlightAll & \textbf{\darkblue{Safety and Efficacy of the BNT162b2 mRNA Covid-19 Vaccine.}} \\ \cline{2-3}
        & Overlapper & \textbf{\darkblue{Safety}} and \textbf{\darkblue{Efficacy}} of the \textbf{\darkblue{BNT162b2 mRNA Covid-19 Vaccine.}} \\ \cline{2-3}
        & BM25 & Safety \textbf{\darkblue{and}} Efficacy of \textbf{\darkblue{the}} BNT162b2 mRNA Covid-19 \textbf{\darkblue{Vaccine}}. \\ \cline{2-3}
        & Word2Vec & \textbf{\darkblue{Safety}} and \textbf{\darkblue{Efficacy}} of the BNT162b2 mRNA Covid-19 \textbf{\darkblue{Vaccine}}. \\ \cline{2-3}
        & BioWord2Vec & \textbf{\darkblue{Safety}} and \textbf{\darkblue{Efficacy}} of the BNT162b2 mRNA Covid-19 \textbf{\darkblue{Vaccine}}. \\ \cline{2-3}
        & MPNet & \textbf{\darkblue{Safety}} and Efficacy of the BNT162b2 mRNA \textbf{\darkblue{Covid-19 Vaccine}}. \\ \cline{2-3}
        & MedCPT & Safety \textbf{\darkblue{and}} Efficacy of the BNT162b2 \textbf{\darkblue{mRNA}} Covid-19 \textbf{\darkblue{Vaccine}} \\ \cline{2-3} 
        & GPT-3.5-Turbo & \textbf{\darkblue{Safety}} and \textbf{\darkblue{Efficacy}} of the \textbf{\darkblue{BNT162b2}} \textbf{\darkblue{mRNA}} Covid-19 Vaccine  \\ \cline{2-3}      
        & GPT-4 & Safety and Efficacy of the \textbf{\darkblue{BNT162b2}} \textbf{\darkblue{mRNA}} \textbf{\darkblue{Covid-19}} \textbf{\darkblue{Vaccine}} \\ \hline \hline

        \multirow{3}{4.5cm}{\newline\newline\newline\newline \textbf{Seed article PMID }: 23651522 \newline\textbf{Similar article PMID}: 29361967 \newline\newline \textbf{Seed article title}: Very-low-carbohydrate ketogenic diet v. low-fat diet for long-term weight loss: a meta-analysis of randomized controlled trials.} & \textbf{Similar article title} & The effect of a ketogenic diet versus a high-carbohydrate, low-fat diet on sleep ... and cardiovascular health independent of weight loss ... \\ \cline{2-3}
        & \textbf{Ground truth} & The effect of a \textbf{\darkblue{ketogenic diet}} versus a high-carbohydrate, \textbf{\darkblue{low-fat diet}} on sleep ... and cardiovascular health independent of \textbf{\darkblue{weight}} loss ... \\ \cline{2-3}
        & \textbf{HSAT (ours)} & The effect of a \textbf{\darkblue{ketogenic diet}} versus a high-carbohydrate, \textbf{\darkblue{low-fat diet}} on sleep ... and cardiovascular health independent of \textbf{\darkblue{weight}} loss ... \\ \cline{2-3}
        & BM25 & The effect of a \textbf{\darkblue{ketogenic diet}} versus a high-carbohydrate, \textbf{\darkblue{low-fat diet}} on sleep ... and cardiovascular health independent of weight loss ... \\ \cline{2-3}
        & Word2Vec & The effect of a \textbf{\darkblue{ketogenic diet}} versus a high-carbohydrate, \textbf{\darkblue{low-fat diet}} on sleep ... and cardiovascular health independent of weight loss ... \\ \cline{2-3}
        & BioWord2Vec & The effect of a ketogenic \textbf{\darkblue{diet}} versus a \textbf{\darkblue{high-carbohydrate}}, low-fat \textbf{\darkblue{diet}} on sleep ... and cardiovascular health independent of weight loss ... \\ \cline{2-3}
        & MPNet & The effect of a \textbf{\darkblue{ketogenic}} diet versus a \textbf{\darkblue{high-carbohydrate}}, \textbf{\darkblue{low-fat}} diet on sleep ... and cardiovascular health independent of weight loss ... \\ \cline{2-3}
        & MedCPT & The effect of a \textbf{\darkblue{ketogenic diet}} versus a high-carbohydrate, \textbf{\darkblue{low-fat diet}} on sleep ... and cardiovascular health independent of weight loss ... \\ \cline{2-3}
        & GPT-3.5-Turbo & The effect of a \textbf{\darkblue{ketogenic diet}} versus a \textbf{\darkblue{high-carbohydrate}}, \textbf{\darkblue{low-fat diet}} on sleep ... and cardiovascular health independent of weight loss ... \\ \cline{2-3}        
        & GPT-4 & The effect of a \textbf{\darkblue{ketogenic diet}} versus a high-carbohydrate, \textbf{\darkblue{low-fat diet}} on sleep ... and cardiovascular health independent of weight loss ... \\ \hline \hline
        
        \multirow{3}{4.5cm}{\newline\newline\newline \textbf{Seed article PMID}: 27533649 \newline \textbf{Similar article PMID}: 25650566 \newline\newline \textbf{Seed article title}: Efficacy of Turmeric Extracts and Curcumin for Alleviating the Symptoms of Joint Arthritis: A Systematic Review and Meta-Analysis of Randomized Clinical Trials.} & \textbf{Similar article title} & Mobile bearing vs fixed bearing prostheses for posterior cruciate retaining total knee arthroplasty ... patients with osteoarthritis and rheumatoid arthritis. \\ \cline{2-3}
        & \textbf{Ground truth} & Mobile bearing vs fixed bearing prostheses for posterior cruciate retaining total knee arthroplasty ... patients with \textbf{\darkblue{osteoarthritis}} and \textbf{\darkblue{rheumatoid arthritis}}. \\ \cline{2-3}
        & \textbf{HSAT (ours)} & Mobile bearing vs fixed bearing prostheses for posterior cruciate retaining total knee arthroplasty ... patients with \textbf{\darkblue{osteoarthritis}} and \textbf{\darkblue{rheumatoid arthritis}}. \\ \cline{2-3}
        & BM25 & Mobile bearing vs fixed bearing prostheses for posterior cruciate retaining \textbf{\darkblue{total}} knee arthroplasty ... patients with \textbf{\darkblue{osteoarthritis}} and rheumatoid \textbf{\darkblue{arthritis}}.  \\ \cline{2-3}
        & Word2Vec & Mobile bearing vs fixed bearing prostheses for posterior cruciate retaining total knee \textbf{\darkblue{arthroplasty}} ... patients with osteoarthritis and \textbf{\darkblue{rheumatoid arthritis}}.  \\ \cline{2-3}
        & BioWord2Vec & \textbf{\darkblue{Mobile bearing vs}} fixed bearing prostheses for posterior cruciate retaining total knee arthroplasty ... patients with osteoarthritis and rheumatoid arthritis.  \\ \cline{2-3}
        & MPNet & Mobile bearing vs fixed bearing prostheses for posterior cruciate retaining total knee arthroplasty ... patients with \textbf{\darkblue{osteoarthritis}} and \textbf{\darkblue{rheumatoid arthritis}}.  \\ \cline{2-3}
        & GPT-3.5-Turbo & Mobile bearing vs fixed bearing prostheses for posterior cruciate retaining \textbf{\darkblue{total}} knee arthroplasty ... patients with \textbf{\darkblue{osteoarthritis}} and \textbf{\darkblue{rheumatoid arthritis}}.  \\ \cline{2-3}      
        & GPT-4 & Mobile bearing vs fixed bearing prostheses for posterior cruciate retaining total knee arthroplasty ... \textbf{\darkblue{patients}} with \textbf{\darkblue{osteoarthritis}} and \textbf{\darkblue{rheumatoid arthritis}}. \\ \hline 
    \end{tabular}
    \caption{Case studies of outputs from Highlight Similar Article Title (HSAT) and baselines. The first two entries are from the PubCLogs test set, and the third one is from the manually annotated set. The leftmost column presents the PMIDs and the titles of the seed articles. In the middle column, rows are labelled as the similar article title, the ground truth, HSAT, and various baselines. In the third column, the similar article titles are shown, and ground truths and predicted outputs are highlighted.}
    \label{tab:case_studies}
\end{table*}

\subsection{Analysis}
We divide the test set of PubCLogs into five subgroups. The subgroups are divided with the total combined clicked counts of the article pair. 
A clear trend is that as that most models, including baseline models and HSAT, tend to perform worse on articles with fewer clicks.
This trend could be explained by the fact that the bottom third contains more relatively obscure topics such as ``autoimmune glial fibrillary acidic protein astrocytopathy,'' whereas the top third contains proportionally more of popular topics such as COVID. 
For example, out of 15k instances each, COVID is mentioned more than 2000 times in the top third and about 900 times in the bottom third. 
We hypothesize that HSAT performs better on article pairs that are more highly clicked because they tend to be more popular topics that are widely read and researched. 
Because there are more articles in those areas, the recommendations are bound to be better. 
And since the pairs are more similar to each other, the quality of both the PubCLogs ground truths and HSAT outputs will be higher. 
To test this, we divide the test set into five subgroups based on semantic similarity. We use MedCPT article encoder to encode the seed and similar articles of each entry in the PubCLogs test set, and use cosine similarity to obtain the pairwise relevance. Based on the $\mathrm{F_1}$ scores of HSAT on each subgroup, we find that HSAT tends to do better when article pairs are more semantically similar, as described in Figure~\ref{fig:distribution_semantic_similarity}.

The performance of HSAT on the manual set is slightly lower at 80.62, compared to 91.72 on the test set. This may be due to the smaller size of the set, which contains 100 article pairs, compared to the PubCLogs test set's 47,000. Also, to form the manual set, we randomly selected article pairs from the top 1000 most highly clicked articles, and chose from the top 20 similar articles actually recommended by PubMed. Because many highly clicked articles contain a higher percentage of types of titles that tend to be uninformative, such as ``Methodology of a systematic review,'' the semantic similarity between the pairs is bound to be lower, which would lead to lower HSAT performance as mentioned above. 

\paragraph{\textbf{Future work}}
Since HSAT training depends on our collected dataset from user coclick logs, the prediction quality for PMIDs that do not get searched as often may not be as good. And since a substantial proportion of PubMed queries are long tail, we could find other ways to complement long tail instances. A source of noise in the training data might be simple article titles that do not provide much insight into their content, such as \textit{Report of the 49th annual meeting of the Pacific Association of Pediatric Surgeons, Kaua'i, Hawai'i, April 24th-28th, 2016} (PMID 27680595). In addition, because every user will have different prior knowledge and information seeking goals, personalized recommendation would be the ultimate goal for tasks such as highlighting similar article titles.  

\section{Conclusion}
We approach the important problem of explaining literature recommendations by highlighting relevant tokens in the title of recommended articles. We utilize user query logs to construct the PubCLogs dataset by pairing coclicked articles into seed and similar articles, and using the respective queries to label relevant tokens in the similar article titles. To demonstrate the value of harnessing user intelligence, we train Highlight Similar Article Title (HSAT), a transformer-based model, on PubCLogs with a sequence tagging objective. HSAT outperforms strong common baselines, including BM25, Word2Vec, MPNet, and GPT-4. Our results on the PubCLogs test set, a separate manually annotated test set, and a user study confirm that user search logs can be effectively repurposed to provide explanations for article recommendations.

\begin{acks}
This research was supported by the Intramural Research Program of the National Library of Medicine (NLM), National Institutes of Health.
\end{acks}

\bibliographystyle{ACM-Reference-Format}
\bibliography{references}


\begin{thebibliography}{43}


\ifx \showCODEN    \undefined \def \showCODEN     #1{\unskip}     \fi
\ifx \showDOI      \undefined \def \showDOI       #1{#1}\fi
\ifx \showISBNx    \undefined \def \showISBNx     #1{\unskip}     \fi
\ifx \showISBNxiii \undefined \def \showISBNxiii  #1{\unskip}     \fi
\ifx \showISSN     \undefined \def \showISSN      #1{\unskip}     \fi
\ifx \showLCCN     \undefined \def \showLCCN      #1{\unskip}     \fi
\ifx \shownote     \undefined \def \shownote      #1{#1}          \fi
\ifx \showarticletitle \undefined \def \showarticletitle #1{#1}   \fi
\ifx \showURL      \undefined \def \showURL       {\relax}        \fi
\providecommand\bibfield[2]{#2}
\providecommand\bibinfo[2]{#2}
\providecommand\natexlab[1]{#1}
\providecommand\showeprint[2][]{arXiv:#2}

\bibitem[Ai and Narayanan.R(2021)]%
        {qingyao2021model-agnostic}
\bibfield{author}{\bibinfo{person}{Qingyao Ai} {and} \bibinfo{person}{Lakshmi Narayanan.R}.} \bibinfo{year}{2021}\natexlab{}.
\newblock \showarticletitle{Model-Agnostic vs. Model-Intrinsic Interpretability for Explainable Product Search}. In \bibinfo{booktitle}{\emph{Proceedings of the 30th ACM International Conference on Information \& Knowledge Management}} (Virtual Event, Queensland, Australia) \emph{(\bibinfo{series}{CIKM '21})}. \bibinfo{publisher}{Association for Computing Machinery}, \bibinfo{address}{New York, NY, USA}, \bibinfo{pages}{5–15}.
\newblock
\showISBNx{9781450384469}
\urldef\tempurl%
\url{https://doi.org/10.1145/3459637.3482276}
\showDOI{\tempurl}


\bibitem[Baumgartner~Jr et~al\mbox{.}(2007)]%
        {baumgartner2007manual}
\bibfield{author}{\bibinfo{person}{William~A Baumgartner~Jr}, \bibinfo{person}{K~Bretonnel Cohen}, \bibinfo{person}{Lynne~M Fox}, \bibinfo{person}{George Acquaah-Mensah}, {and} \bibinfo{person}{Lawrence Hunter}.} \bibinfo{year}{2007}\natexlab{}.
\newblock \showarticletitle{Manual curation is not sufficient for annotation of genomic databases}.
\newblock \bibinfo{journal}{\emph{Bioinformatics}} \bibinfo{volume}{23}, \bibinfo{number}{13} (\bibinfo{year}{2007}), \bibinfo{pages}{i41--i48}.
\newblock


\bibitem[Bioinformatics et~al\mbox{.}(2007)]%
        {Lin2007PubMedRA}
\bibfield{author}{\bibinfo{person}{Bmc Bioinformatics}, \bibinfo{person}{Jimmy~J. Lin}, \bibinfo{person}{John Wilbur}, {and} \bibinfo{person}{Email}.} \bibinfo{year}{2007}\natexlab{}.
\newblock \showarticletitle{PubMed related articles: a probabilistic topic-based model for content similarity}.
\newblock \bibinfo{journal}{\emph{BMC Bioinformatics}}  \bibinfo{volume}{8} (\bibinfo{year}{2007}), \bibinfo{pages}{423 -- 423}.
\newblock
\urldef\tempurl%
\url{https://api.semanticscholar.org/CorpusID:3201001}
\showURL{%
\tempurl}


\bibitem[Bird et~al\mbox{.}(2009)]%
        {nltk2009}
\bibfield{author}{\bibinfo{person}{Steven Bird}, \bibinfo{person}{Ewan Klein}, {and} \bibinfo{person}{Edward Loper}.} \bibinfo{year}{2009}\natexlab{}.
\newblock \bibinfo{booktitle}{\emph{Natural Language Processing with Python} (\bibinfo{edition}{1st} ed.)}.
\newblock \bibinfo{publisher}{O'Reilly Media, Inc.}
\newblock
\showISBNx{0596516495}


\bibitem[Cachola et~al\mbox{.}(2020)]%
        {cachola2020TLDR}
\bibfield{author}{\bibinfo{person}{Isabel Cachola}, \bibinfo{person}{Kyle Lo}, \bibinfo{person}{Arman Cohan}, {and} \bibinfo{person}{Daniel~S. Weld}.} \bibinfo{year}{2020}\natexlab{}.
\newblock \showarticletitle{TLDR: Extreme Summarization of Scientific Documents}.
\newblock \bibinfo{journal}{\emph{ArXiv}}  \bibinfo{volume}{abs/2004.15011} (\bibinfo{year}{2020}).
\newblock
\urldef\tempurl%
\url{https://api.semanticscholar.org/CorpusID:216867622}
\showURL{%
\tempurl}


\bibitem[Choi et~al\mbox{.}(2018)]%
        {QuAC}
\bibfield{author}{\bibinfo{person}{Eunsol Choi}, \bibinfo{person}{He He}, \bibinfo{person}{Mohit Iyyer}, \bibinfo{person}{Mark Yatskar}, \bibinfo{person}{Wen tau Yih}, \bibinfo{person}{Yejin Choi}, \bibinfo{person}{Percy Liang}, {and} \bibinfo{person}{Luke Zettlemoyer}.} \bibinfo{year}{2018}\natexlab{}.
\newblock \showarticletitle{QuAC: Question Answering in Context}. In \bibinfo{booktitle}{\emph{Conference on Empirical Methods in Natural Language Processing}}.
\newblock
\urldef\tempurl%
\url{https://api.semanticscholar.org/CorpusID:52057510}
\showURL{%
\tempurl}


\bibitem[Dasigi et~al\mbox{.}(2019)]%
        {dasigi2019quoref}
\bibfield{author}{\bibinfo{person}{Pradeep Dasigi}, \bibinfo{person}{Nelson~F. Liu}, \bibinfo{person}{Ana Marasovi{\'c}}, \bibinfo{person}{Noah~A. Smith}, {and} \bibinfo{person}{Matt Gardner}.} \bibinfo{year}{2019}\natexlab{}.
\newblock \showarticletitle{{Q}uoref: A Reading Comprehension Dataset with Questions Requiring Coreferential Reasoning}. In \bibinfo{booktitle}{\emph{Proceedings of the 2019 Conference on Empirical Methods in Natural Language Processing and the 9th International Joint Conference on Natural Language Processing (EMNLP-IJCNLP)}}, \bibfield{editor}{\bibinfo{person}{Kentaro Inui}, \bibinfo{person}{Jing Jiang}, \bibinfo{person}{Vincent Ng}, {and} \bibinfo{person}{Xiaojun Wan}} (Eds.). \bibinfo{publisher}{Association for Computational Linguistics}, \bibinfo{address}{Hong Kong, China}, \bibinfo{pages}{5925--5932}.
\newblock
\urldef\tempurl%
\url{https://doi.org/10.18653/v1/D19-1606}
\showDOI{\tempurl}


\bibitem[Devlin et~al\mbox{.}(2019)]%
        {bert}
\bibfield{author}{\bibinfo{person}{Jacob Devlin}, \bibinfo{person}{Ming-Wei Chang}, \bibinfo{person}{Kenton Lee}, {and} \bibinfo{person}{Kristina Toutanova}.} \bibinfo{year}{2019}\natexlab{}.
\newblock \showarticletitle{{BERT}: Pre-training of Deep Bidirectional Transformers for Language Understanding}. In \bibinfo{booktitle}{\emph{Proceedings of the 2019 Conference of the North {A}merican Chapter of the Association for Computational Linguistics: Human Language Technologies, Volume 1 (Long and Short Papers)}}, \bibfield{editor}{\bibinfo{person}{Jill Burstein}, \bibinfo{person}{Christy Doran}, {and} \bibinfo{person}{Thamar Solorio}} (Eds.). \bibinfo{publisher}{Association for Computational Linguistics}, \bibinfo{address}{Minneapolis, Minnesota}, \bibinfo{pages}{4171--4186}.
\newblock
\urldef\tempurl%
\url{https://doi.org/10.18653/v1/N19-1423}
\showDOI{\tempurl}


\bibitem[Dogan et~al\mbox{.}(2009)]%
        {Dogan2009}
\bibfield{author}{\bibinfo{person}{Rezarta~Islamaj Dogan}, \bibinfo{person}{G.~Craig Murray}, \bibinfo{person}{Aur{\'e}lie N{\'e}v{\'e}ol}, {and} \bibinfo{person}{Zhiyong Lu}.} \bibinfo{year}{2009}\natexlab{}.
\newblock \showarticletitle{Understanding PubMed{\textregistered} user search behavior through log analysis}.
\newblock \bibinfo{journal}{\emph{Database: The Journal of Biological Databases and Curation}}  \bibinfo{volume}{2009} (\bibinfo{year}{2009}).
\newblock


\bibitem[Dua et~al\mbox{.}(2019)]%
        {dua2019drop}
\bibfield{author}{\bibinfo{person}{Dheeru Dua}, \bibinfo{person}{Yizhong Wang}, \bibinfo{person}{Pradeep Dasigi}, \bibinfo{person}{Gabriel Stanovsky}, \bibinfo{person}{Sameer Singh}, {and} \bibinfo{person}{Matt Gardner}.} \bibinfo{year}{2019}\natexlab{}.
\newblock \showarticletitle{{DROP}: A Reading Comprehension Benchmark Requiring Discrete Reasoning Over Paragraphs}. In \bibinfo{booktitle}{\emph{Proceedings of the 2019 Conference of the North {A}merican Chapter of the Association for Computational Linguistics: Human Language Technologies, Volume 1 (Long and Short Papers)}}, \bibfield{editor}{\bibinfo{person}{Jill Burstein}, \bibinfo{person}{Christy Doran}, {and} \bibinfo{person}{Thamar Solorio}} (Eds.). \bibinfo{publisher}{Association for Computational Linguistics}, \bibinfo{address}{Minneapolis, Minnesota}, \bibinfo{pages}{2368--2378}.
\newblock
\urldef\tempurl%
\url{https://doi.org/10.18653/v1/N19-1246}
\showDOI{\tempurl}


\bibitem[Ely et~al\mbox{.}(2005)]%
        {ely2005answering}
\bibfield{author}{\bibinfo{person}{John~W Ely}, \bibinfo{person}{Jerome~A Osheroff}, \bibinfo{person}{M~Lee Chambliss}, \bibinfo{person}{Mark~H Ebell}, {and} \bibinfo{person}{Marcy~E Rosenbaum}.} \bibinfo{year}{2005}\natexlab{}.
\newblock \showarticletitle{Answering physicians' clinical questions: obstacles and potential solutions}.
\newblock \bibinfo{journal}{\emph{Journal of the American Medical Informatics Association}} \bibinfo{volume}{12}, \bibinfo{number}{2} (\bibinfo{year}{2005}), \bibinfo{pages}{217--224}.
\newblock


\bibitem[Fiorini et~al\mbox{.}(2018)]%
        {Fiorini2018}
\bibfield{author}{\bibinfo{person}{Nicolas Fiorini}, \bibinfo{person}{Robert Leaman}, \bibinfo{person}{David~J. Lipman}, {and} \bibinfo{person}{Zhiyong Lu}.} \bibinfo{year}{2018}\natexlab{}.
\newblock \showarticletitle{How user intelligence is improving PubMed}.
\newblock \bibinfo{journal}{\emph{Nature Biotechnology}}  \bibinfo{volume}{36} (\bibinfo{year}{2018}), \bibinfo{pages}{937--945}.
\newblock
\urldef\tempurl%
\url{https://api.semanticscholar.org/CorpusID:52892576}
\showURL{%
\tempurl}


\bibitem[Gopalakrishnan et~al\mbox{.}(2019)]%
        {gopalakrishnan2019survey}
\bibfield{author}{\bibinfo{person}{Vishrawas Gopalakrishnan}, \bibinfo{person}{Kishlay Jha}, \bibinfo{person}{Wei Jin}, {and} \bibinfo{person}{Aidong Zhang}.} \bibinfo{year}{2019}\natexlab{}.
\newblock \showarticletitle{A survey on literature based discovery approaches in biomedical domain}.
\newblock \bibinfo{journal}{\emph{Journal of biomedical informatics}}  \bibinfo{volume}{93} (\bibinfo{year}{2019}), \bibinfo{pages}{103141}.
\newblock


\bibitem[Gu et~al\mbox{.}(2021)]%
        {pubmedbert}
\bibfield{author}{\bibinfo{person}{Yu Gu}, \bibinfo{person}{Robert Tinn}, \bibinfo{person}{Hao Cheng}, \bibinfo{person}{Michael Lucas}, \bibinfo{person}{Naoto Usuyama}, \bibinfo{person}{Xiaodong Liu}, \bibinfo{person}{Tristan Naumann}, \bibinfo{person}{Jianfeng Gao}, {and} \bibinfo{person}{Hoifung Poon}.} \bibinfo{year}{2021}\natexlab{}.
\newblock \showarticletitle{Domain-specific language model pretraining for biomedical natural language processing}.
\newblock \bibinfo{journal}{\emph{ACM Transactions on Computing for Healthcare (HEALTH)}} \bibinfo{volume}{3}, \bibinfo{number}{1} (\bibinfo{year}{2021}), \bibinfo{pages}{1--23}.
\newblock


\bibitem[Jain and Wallace(2019)]%
        {jain2019attention}
\bibfield{author}{\bibinfo{person}{Sarthak Jain} {and} \bibinfo{person}{Byron~C. Wallace}.} \bibinfo{year}{2019}\natexlab{}.
\newblock \showarticletitle{{A}ttention is not {E}xplanation}. In \bibinfo{booktitle}{\emph{Proceedings of the 2019 Conference of the North {A}merican Chapter of the Association for Computational Linguistics: Human Language Technologies, Volume 1 (Long and Short Papers)}}, \bibfield{editor}{\bibinfo{person}{Jill Burstein}, \bibinfo{person}{Christy Doran}, {and} \bibinfo{person}{Thamar Solorio}} (Eds.). \bibinfo{publisher}{Association for Computational Linguistics}, \bibinfo{address}{Minneapolis, Minnesota}, \bibinfo{pages}{3543--3556}.
\newblock
\urldef\tempurl%
\url{https://doi.org/10.18653/v1/N19-1357}
\showDOI{\tempurl}


\bibitem[Jin et~al\mbox{.}(2023a)]%
        {medcpt}
\bibfield{author}{\bibinfo{person}{Qiao Jin}, \bibinfo{person}{Won Kim}, \bibinfo{person}{Qingyu Chen}, \bibinfo{person}{Donald~C Comeau}, \bibinfo{person}{Lana Yeganova}, \bibinfo{person}{W~John Wilbur}, {and} \bibinfo{person}{Zhiyong Lu}.} \bibinfo{year}{2023}\natexlab{a}.
\newblock \showarticletitle{MedCPT: Contrastive Pre-trained Transformers with large-scale PubMed search logs for zero-shot biomedical information retrieval}.
\newblock \bibinfo{journal}{\emph{Bioinformatics}} \bibinfo{volume}{39}, \bibinfo{number}{11} (\bibinfo{year}{2023}), \bibinfo{pages}{btad651}.
\newblock


\bibitem[Jin et~al\mbox{.}(2023b)]%
        {jin2023pubmed}
\bibfield{author}{\bibinfo{person}{Qiao Jin}, \bibinfo{person}{Robert Leaman}, {and} \bibinfo{person}{Zhiyong Lu}.} \bibinfo{year}{2023}\natexlab{b}.
\newblock \bibinfo{title}{PubMed and Beyond: Biomedical Literature Search in the Age of Artificial Intelligence}.
\newblock
\newblock
\showeprint[arxiv]{2307.09683}~[cs.IR]


\bibitem[Jin et~al\mbox{.}(2022)]%
        {jin2022state}
\bibfield{author}{\bibinfo{person}{Qiao Jin}, \bibinfo{person}{Chuanqi Tan}, \bibinfo{person}{Mosha Chen}, \bibinfo{person}{Ming Yan}, \bibinfo{person}{Ningyu Zhang}, \bibinfo{person}{Songfang Huang}, \bibinfo{person}{Xiaozhong Liu}, {et~al\mbox{.}}} \bibinfo{year}{2022}\natexlab{}.
\newblock \showarticletitle{State-of-the-Art Evidence Retriever for Precision Medicine: Algorithm Development and Validation}.
\newblock \bibinfo{journal}{\emph{JMIR Medical Informatics}} \bibinfo{volume}{10}, \bibinfo{number}{12} (\bibinfo{year}{2022}), \bibinfo{pages}{e40743}.
\newblock


\bibitem[Kingma and Ba(2014)]%
        {adam}
\bibfield{author}{\bibinfo{person}{Diederik~P. Kingma} {and} \bibinfo{person}{Jimmy Ba}.} \bibinfo{year}{2014}\natexlab{}.
\newblock \showarticletitle{Adam: A Method for Stochastic Optimization}.
\newblock \bibinfo{journal}{\emph{CoRR}}  \bibinfo{volume}{abs/1412.6980} (\bibinfo{year}{2014}).
\newblock


\bibitem[Li et~al\mbox{.}(2022)]%
        {MultiSpanQA}
\bibfield{author}{\bibinfo{person}{Haonan Li}, \bibinfo{person}{Martin Tomko}, \bibinfo{person}{Maria Vasardani}, {and} \bibinfo{person}{Timothy Baldwin}.} \bibinfo{year}{2022}\natexlab{}.
\newblock \showarticletitle{{M}ulti{S}pan{QA}: A Dataset for Multi-Span Question Answering}. In \bibinfo{booktitle}{\emph{Proceedings of the 2022 Conference of the North American Chapter of the Association for Computational Linguistics: Human Language Technologies}}, \bibfield{editor}{\bibinfo{person}{Marine Carpuat}, \bibinfo{person}{Marie-Catherine de~Marneffe}, {and} \bibinfo{person}{Ivan~Vladimir Meza~Ruiz}} (Eds.). \bibinfo{publisher}{Association for Computational Linguistics}, \bibinfo{address}{Seattle, United States}, \bibinfo{pages}{1250--1260}.
\newblock
\urldef\tempurl%
\url{https://doi.org/10.18653/v1/2022.naacl-main.90}
\showDOI{\tempurl}


\bibitem[Lin(2019)]%
        {Lin2019Hype}
\bibfield{author}{\bibinfo{person}{Jimmy Lin}.} \bibinfo{year}{2019}\natexlab{}.
\newblock \showarticletitle{The Neural Hype and Comparisons Against Weak Baselines}.
\newblock \bibinfo{journal}{\emph{SIGIR Forum}} \bibinfo{volume}{52}, \bibinfo{number}{2} (\bibinfo{date}{jan} \bibinfo{year}{2019}), \bibinfo{pages}{40–51}.
\newblock
\showISSN{0163-5840}
\urldef\tempurl%
\url{https://doi.org/10.1145/3308774.3308781}
\showDOI{\tempurl}


\bibitem[Lin and Wilbur(2009)]%
        {Lin2009modeling}
\bibfield{author}{\bibinfo{person}{Jimmy Lin} {and} \bibinfo{person}{W~John Wilbur}.} \bibinfo{year}{2009}\natexlab{}.
\newblock \showarticletitle{Modeling actions of PubMed users with n-gram language models}.
\newblock \bibinfo{journal}{\emph{Information retrieval}}  \bibinfo{volume}{12} (\bibinfo{year}{2009}), \bibinfo{pages}{487--503}.
\newblock


\bibitem[Mikolov et~al\mbox{.}(2013)]%
        {word2vec}
\bibfield{author}{\bibinfo{person}{Tomas Mikolov}, \bibinfo{person}{Ilya Sutskever}, \bibinfo{person}{Kai Chen}, \bibinfo{person}{Gregory~S. Corrado}, {and} \bibinfo{person}{Jeffrey Dean}.} \bibinfo{year}{2013}\natexlab{}.
\newblock \showarticletitle{Distributed Representations of Words and Phrases and their Compositionality}. In \bibinfo{booktitle}{\emph{Neural Information Processing Systems}}.
\newblock
\urldef\tempurl%
\url{https://api.semanticscholar.org/CorpusID:16447573}
\showURL{%
\tempurl}


\bibitem[OpenAI(2022)]%
        {chatgpt}
\bibfield{author}{\bibinfo{person}{OpenAI}.} \bibinfo{year}{2022}\natexlab{}.
\newblock \bibinfo{title}{Introducing ChatGPT}.
\newblock
\newblock
\newblock
\shownote{\url{https://openai.com/blog/chatgpt}, Last accessed on 2024-01-11}.


\bibitem[OpenAI(2023)]%
        {gpt4}
\bibfield{author}{\bibinfo{person}{OpenAI}.} \bibinfo{year}{2023}\natexlab{}.
\newblock \bibinfo{title}{GPT-4 Technical Report}.
\newblock
\newblock
\showeprint[arxiv]{2303.08774}~[cs.CL]


\bibitem[Paszke et~al\mbox{.}(2019)]%
        {pytorch}
\bibfield{author}{\bibinfo{person}{Adam Paszke}, \bibinfo{person}{Sam Gross}, \bibinfo{person}{Francisco Massa}, \bibinfo{person}{Adam Lerer}, \bibinfo{person}{James Bradbury}, \bibinfo{person}{Gregory Chanan}, \bibinfo{person}{Trevor Killeen}, \bibinfo{person}{Zeming Lin}, \bibinfo{person}{Natalia Gimelshein}, \bibinfo{person}{Luca Antiga}, \bibinfo{person}{Alban Desmaison}, \bibinfo{person}{Andreas K{\"{o}}pf}, \bibinfo{person}{Edward~Z. Yang}, \bibinfo{person}{Zach DeVito}, \bibinfo{person}{Martin Raison}, \bibinfo{person}{Alykhan Tejani}, \bibinfo{person}{Sasank Chilamkurthy}, \bibinfo{person}{Benoit Steiner}, \bibinfo{person}{Lu Fang}, \bibinfo{person}{Junjie Bai}, {and} \bibinfo{person}{Soumith Chintala}.} \bibinfo{year}{2019}\natexlab{}.
\newblock \showarticletitle{PyTorch: An Imperative Style, High-Performance Deep Learning Library}.
\newblock \bibinfo{journal}{\emph{CoRR}}  \bibinfo{volume}{abs/1912.01703} (\bibinfo{year}{2019}).
\newblock
\showeprint[arXiv]{1912.01703}
\urldef\tempurl%
\url{http://arxiv.org/abs/1912.01703}
\showURL{%
\tempurl}


\bibitem[Peake and Wang(2018)]%
        {peake2018posthoc}
\bibfield{author}{\bibinfo{person}{Georgina Peake} {and} \bibinfo{person}{Jun Wang}.} \bibinfo{year}{2018}\natexlab{}.
\newblock \showarticletitle{Explanation mining: Post hoc interpretability of latent factor models for recommendation systems}. In \bibinfo{booktitle}{\emph{Proceedings of the 24th ACM SIGKDD International Conference on Knowledge Discovery \& Data Mining}}. \bibinfo{pages}{2060--2069}.
\newblock


\bibitem[Rajpurkar et~al\mbox{.}(2018)]%
        {SQuAD2.0}
\bibfield{author}{\bibinfo{person}{Pranav Rajpurkar}, \bibinfo{person}{Robin Jia}, {and} \bibinfo{person}{Percy Liang}.} \bibinfo{year}{2018}\natexlab{}.
\newblock \showarticletitle{Know What You Don’t Know: Unanswerable Questions for SQuAD}.
\newblock \bibinfo{journal}{\emph{ArXiv}}  \bibinfo{volume}{abs/1806.03822} (\bibinfo{year}{2018}).
\newblock
\urldef\tempurl%
\url{https://api.semanticscholar.org/CorpusID:47018994}
\showURL{%
\tempurl}


\bibitem[Rajpurkar et~al\mbox{.}(2016)]%
        {SQuAD}
\bibfield{author}{\bibinfo{person}{Pranav Rajpurkar}, \bibinfo{person}{Jian Zhang}, \bibinfo{person}{Konstantin Lopyrev}, {and} \bibinfo{person}{Percy Liang}.} \bibinfo{year}{2016}\natexlab{}.
\newblock \showarticletitle{SQuAD: 100,000+ Questions for Machine Comprehension of Text}. In \bibinfo{booktitle}{\emph{Conference on Empirical Methods in Natural Language Processing}}.
\newblock
\urldef\tempurl%
\url{https://api.semanticscholar.org/CorpusID:11816014}
\showURL{%
\tempurl}


\bibitem[Raza et~al\mbox{.}(2022)]%
        {raza2022large}
\bibfield{author}{\bibinfo{person}{Shaina Raza}, \bibinfo{person}{Deepak~John Reji}, \bibinfo{person}{Femi Shajan}, {and} \bibinfo{person}{Syed~Raza Bashir}.} \bibinfo{year}{2022}\natexlab{}.
\newblock \showarticletitle{Large-scale application of named entity recognition to biomedicine and epidemiology}.
\newblock \bibinfo{journal}{\emph{PLOS Digital Health}} \bibinfo{volume}{1}, \bibinfo{number}{12} (\bibinfo{year}{2022}), \bibinfo{pages}{e0000152}.
\newblock


\bibitem[Robertson and Zaragoza(2009)]%
        {bm25}
\bibfield{author}{\bibinfo{person}{Stephen~E. Robertson} {and} \bibinfo{person}{Hugo Zaragoza}.} \bibinfo{year}{2009}\natexlab{}.
\newblock \showarticletitle{The Probabilistic Relevance Framework: {BM25} and Beyond}.
\newblock \bibinfo{journal}{\emph{Found. Trends Inf. Retr.}} \bibinfo{volume}{3}, \bibinfo{number}{4} (\bibinfo{year}{2009}), \bibinfo{pages}{333--389}.
\newblock
\urldef\tempurl%
\url{https://doi.org/10.1561/1500000019}
\showDOI{\tempurl}


\bibitem[Sayers(2018)]%
        {EntrezUtilities}
\bibfield{author}{\bibinfo{person}{Eric Sayers}.} \bibinfo{year}{2018}\natexlab{}.
\newblock \bibinfo{title}{Entrez Programming Utilities Help}.
\newblock \bibinfo{howpublished}{\url{https://www.ncbi.nlm.nih.gov/books/NBK25501/}}.
\newblock
\newblock
\shownote{National Center for Biotechnology Information (NCBI)}.


\bibitem[Segal et~al\mbox{.}(2020)]%
        {segal-2020-simple}
\bibfield{author}{\bibinfo{person}{Elad Segal}, \bibinfo{person}{Avia Efrat}, \bibinfo{person}{Mor Shoham}, \bibinfo{person}{Amir Globerson}, {and} \bibinfo{person}{Jonathan Berant}.} \bibinfo{year}{2020}\natexlab{}.
\newblock \showarticletitle{A Simple and Effective Model for Answering Multi-span Questions}. In \bibinfo{booktitle}{\emph{Proceedings of the 2020 Conference on Empirical Methods in Natural Language Processing (EMNLP)}}, \bibfield{editor}{\bibinfo{person}{Bonnie Webber}, \bibinfo{person}{Trevor Cohn}, \bibinfo{person}{Yulan He}, {and} \bibinfo{person}{Yang Liu}} (Eds.). \bibinfo{publisher}{Association for Computational Linguistics}, \bibinfo{pages}{3074--3080}.
\newblock
\urldef\tempurl%
\url{https://doi.org/10.18653/v1/2020.emnlp-main.248}
\showDOI{\tempurl}


\bibitem[Singh and Anand(2019)]%
        {singh2019explainable}
\bibfield{author}{\bibinfo{person}{Jaspreet Singh} {and} \bibinfo{person}{Avishek Anand}.} \bibinfo{year}{2019}\natexlab{}.
\newblock \showarticletitle{EXS: Explainable Search Using Local Model Agnostic Interpretability}. In \bibinfo{booktitle}{\emph{Proceedings of the Twelfth ACM International Conference on Web Search and Data Mining}} (Melbourne VIC, Australia) \emph{(\bibinfo{series}{WSDM '19})}. \bibinfo{publisher}{Association for Computing Machinery}, \bibinfo{address}{New York, NY, USA}, \bibinfo{pages}{770–773}.
\newblock
\showISBNx{9781450359405}
\urldef\tempurl%
\url{https://doi.org/10.1145/3289600.3290620}
\showDOI{\tempurl}


\bibitem[Song et~al\mbox{.}(2020)]%
        {mpnet}
\bibfield{author}{\bibinfo{person}{Kaitao Song}, \bibinfo{person}{Xu Tan}, \bibinfo{person}{Tao Qin}, \bibinfo{person}{Jianfeng Lu}, {and} \bibinfo{person}{Tie-Yan Liu}.} \bibinfo{year}{2020}\natexlab{}.
\newblock \showarticletitle{MPNet: Masked and Permuted Pre-training for Language Understanding}. In \bibinfo{booktitle}{\emph{Advances in Neural Information Processing Systems}}, \bibfield{editor}{\bibinfo{person}{H.~Larochelle}, \bibinfo{person}{M.~Ranzato}, \bibinfo{person}{R.~Hadsell}, \bibinfo{person}{M.F. Balcan}, {and} \bibinfo{person}{H.~Lin}} (Eds.), Vol.~\bibinfo{volume}{33}. \bibinfo{publisher}{Curran Associates, Inc.}, \bibinfo{pages}{16857--16867}.
\newblock
\urldef\tempurl%
\url{https://proceedings.neurips.cc/paper_files/paper/2020/file/c3a690be93aa602ee2dc0ccab5b7b67e-Paper.pdf}
\showURL{%
\tempurl}


\bibitem[Vaswani et~al\mbox{.}(2017)]%
        {transformer}
\bibfield{author}{\bibinfo{person}{Ashish Vaswani}, \bibinfo{person}{Noam Shazeer}, \bibinfo{person}{Niki Parmar}, \bibinfo{person}{Jakob Uszkoreit}, \bibinfo{person}{Llion Jones}, \bibinfo{person}{Aidan~N Gomez}, \bibinfo{person}{\L~ukasz Kaiser}, {and} \bibinfo{person}{Illia Polosukhin}.} \bibinfo{year}{2017}\natexlab{}.
\newblock \showarticletitle{Attention is All you Need}. In \bibinfo{booktitle}{\emph{Advances in Neural Information Processing Systems}}, \bibfield{editor}{\bibinfo{person}{I.~Guyon}, \bibinfo{person}{U.~Von Luxburg}, \bibinfo{person}{S.~Bengio}, \bibinfo{person}{H.~Wallach}, \bibinfo{person}{R.~Fergus}, \bibinfo{person}{S.~Vishwanathan}, {and} \bibinfo{person}{R.~Garnett}} (Eds.), Vol.~\bibinfo{volume}{30}. \bibinfo{publisher}{Curran Associates, Inc.}
\newblock
\urldef\tempurl%
\url{https://proceedings.neurips.cc/paper_files/paper/2017/file/3f5ee243547dee91fbd053c1c4a845aa-Paper.pdf}
\showURL{%
\tempurl}


\bibitem[Wang et~al\mbox{.}(2023)]%
        {guido2023screening}
\bibfield{author}{\bibinfo{person}{Shuai Wang}, \bibinfo{person}{Harrisen Scells}, \bibinfo{person}{Bevan Koopman}, {and} \bibinfo{person}{Guido Zuccon}.} \bibinfo{year}{2023}\natexlab{}.
\newblock \showarticletitle{Neural Rankers for Effective Screening Prioritisation in Medical Systematic Review Literature Search}. In \bibinfo{booktitle}{\emph{Proceedings of the 26th Australasian Document Computing Symposium}} (Adelaide, Australia) \emph{(\bibinfo{series}{ADCS '22})}. \bibinfo{publisher}{Association for Computing Machinery}, \bibinfo{address}{New York, NY, USA}, Article \bibinfo{articleno}{4}, \bibinfo{numpages}{10}~pages.
\newblock
\showISBNx{9798400700217}
\urldef\tempurl%
\url{https://doi.org/10.1145/3572960.3572980}
\showDOI{\tempurl}


\bibitem[Wolf et~al\mbox{.}(2020)]%
        {huggingface}
\bibfield{author}{\bibinfo{person}{Thomas Wolf}, \bibinfo{person}{Lysandre Debut}, \bibinfo{person}{Victor Sanh}, \bibinfo{person}{Julien Chaumond}, \bibinfo{person}{Clement Delangue}, \bibinfo{person}{Anthony Moi}, \bibinfo{person}{Pierric Cistac}, \bibinfo{person}{Tim Rault}, \bibinfo{person}{Remi Louf}, \bibinfo{person}{Morgan Funtowicz}, \bibinfo{person}{Joe Davison}, \bibinfo{person}{Sam Shleifer}, \bibinfo{person}{Patrick von Platen}, \bibinfo{person}{Clara Ma}, \bibinfo{person}{Yacine Jernite}, \bibinfo{person}{Julien Plu}, \bibinfo{person}{Canwen Xu}, \bibinfo{person}{Teven Le~Scao}, \bibinfo{person}{Sylvain Gugger}, \bibinfo{person}{Mariama Drame}, \bibinfo{person}{Quentin Lhoest}, {and} \bibinfo{person}{Alexander Rush}.} \bibinfo{year}{2020}\natexlab{}.
\newblock \showarticletitle{Transformers: State-of-the-Art Natural Language Processing}. In \bibinfo{booktitle}{\emph{Proceedings of the 2020 Conference on Empirical Methods in Natural Language Processing: System Demonstrations}}. \bibinfo{publisher}{Association for Computational Linguistics}, \bibinfo{address}{Online}, \bibinfo{pages}{38--45}.
\newblock
\urldef\tempurl%
\url{https://doi.org/10.18653/v1/2020.emnlp-demos.6}
\showDOI{\tempurl}


\bibitem[Yang et~al\mbox{.}(2019)]%
        {xlnet}
\bibfield{author}{\bibinfo{person}{Zhilin Yang}, \bibinfo{person}{Zihang Dai}, \bibinfo{person}{Yiming Yang}, \bibinfo{person}{Jaime Carbonell}, \bibinfo{person}{Ruslan Salakhutdinov}, {and} \bibinfo{person}{Quoc~V. Le}.} \bibinfo{year}{2019}\natexlab{}.
\newblock \bibinfo{booktitle}{\emph{XLNet: generalized autoregressive pretraining for language understanding}}.
\newblock


\bibitem[Yang et~al\mbox{.}(2018)]%
        {HotpotQA}
\bibfield{author}{\bibinfo{person}{Zhilin Yang}, \bibinfo{person}{Peng Qi}, \bibinfo{person}{Saizheng Zhang}, \bibinfo{person}{Yoshua Bengio}, \bibinfo{person}{William~W. Cohen}, \bibinfo{person}{Ruslan Salakhutdinov}, {and} \bibinfo{person}{Christopher~D. Manning}.} \bibinfo{year}{2018}\natexlab{}.
\newblock \showarticletitle{HotpotQA: A Dataset for Diverse, Explainable Multi-hop Question Answering}. In \bibinfo{booktitle}{\emph{Conference on Empirical Methods in Natural Language Processing}}.
\newblock
\urldef\tempurl%
\url{https://api.semanticscholar.org/CorpusID:52822214}
\showURL{%
\tempurl}


\bibitem[Zhang et~al\mbox{.}(2019)]%
        {bioword2vec}
\bibfield{author}{\bibinfo{person}{Yijia Zhang}, \bibinfo{person}{Qingyu Chen}, \bibinfo{person}{Zhihao Yang}, \bibinfo{person}{Hongfei Lin}, {and} \bibinfo{person}{Zhiyong Lu}.} \bibinfo{year}{2019}\natexlab{}.
\newblock \showarticletitle{BioWordVec, improving biomedical word embeddings with subword information and MeSH}.
\newblock \bibinfo{journal}{\emph{Scientific Data}}  \bibinfo{volume}{6} (\bibinfo{year}{2019}).
\newblock
\urldef\tempurl%
\url{https://api.semanticscholar.org/CorpusID:149445302}
\showURL{%
\tempurl}


\bibitem[Zhang et~al\mbox{.}(2020)]%
        {zhang2020explainable}
\bibfield{author}{\bibinfo{person}{Yongfeng Zhang}, \bibinfo{person}{Xu Chen}, {et~al\mbox{.}}} \bibinfo{year}{2020}\natexlab{}.
\newblock \showarticletitle{Explainable recommendation: A survey and new perspectives}.
\newblock \bibinfo{journal}{\emph{Foundations and Trends{\textregistered} in Information Retrieval}} \bibinfo{volume}{14}, \bibinfo{number}{1} (\bibinfo{year}{2020}), \bibinfo{pages}{1--101}.
\newblock


\bibitem[Zhang et~al\mbox{.}(2014)]%
        {zhang2014explicit}
\bibfield{author}{\bibinfo{person}{Yongfeng Zhang}, \bibinfo{person}{Guokun Lai}, \bibinfo{person}{Min Zhang}, \bibinfo{person}{Yi Zhang}, \bibinfo{person}{Yiqun Liu}, {and} \bibinfo{person}{Shaoping Ma}.} \bibinfo{year}{2014}\natexlab{}.
\newblock \showarticletitle{Explicit factor models for explainable recommendation based on phrase-level sentiment analysis}. In \bibinfo{booktitle}{\emph{Proceedings of the 37th international ACM SIGIR conference on Research \& development in information retrieval}}. \bibinfo{pages}{83--92}.
\newblock


\end{thebibliography}


\end{document}